\documentclass[prb,twocolumn,amsmath,amssymb,floatfix,aps,showkeys,showpacs]{revtex4}
\usepackage{graphicx}
\usepackage{dcolumn}
\usepackage{bm}

\usepackage[normalem]{ulem}

\begin{document}

\title{Excitations in high-dimensional random-field Ising magnets}

\author{Bj\"orn Ahrens}
\author{Alexander K. Hartmann}
\email{a.hartmann@uni-oldenburg.de}
\affiliation{Institute of Physics, University of Oldenburg, 26111 Oldenburg, 
Germany}
\date{\today}
%
\begin{abstract}
Domain walls and droplet-like excitation of the random-field Ising
magnet are studied in $d=\{3,4,5,6,7\}$ dimensions by means of exact
numerical ground-state calculations. They are obtained using the
established mapping to the graph-theoretical maximum-flow
problem. This allows to study large system sizes of more than five
million spins in exact thermal equilibrium. All simulations are
carried out at the critical point for the strength $h$ of the
random fields , $h=h_c(d)$, respectively. Using
finite-size scaling, energetic and geometric properties like
stiffness exponents and fractal dimensions  are calculated.
Using these results, we test (hyper) scaling relations, which seem
to be fulfilled below the upper critical dimension $d_u=6$.
Also, for $d<d_u$, the stiffness exponent can be obtained from
the scaling of the ground-state energy.
\end{abstract}


\maketitle
\section{\label{Introduction}Introduction}
The random-field Ising magnet (RFIM) is one of the most frequently 
studied models for magnetic systems with quenched disorder. 
For $d = 3$ and higher 
dimensions, \cite{bricmont1987} it is known to undergo a  
phase transition 
\cite{GofmanAdlerAharonyHarrisSchwartz1993,Rieger1995,Nowak1998,art_uli1999,
HartmannYoung2001,middleton2002,frontera2002,seppala2002,Hartmann2002,
middleton2002b,fes_rfim2008,AhrensHartmann2011} at a critical temperature $T_c$
which
depends on the disorder strength. 
For low temperatures and weak disorder  the ferromagnetic interactions 
dominate and the system is ferromagnetically long-range ordered. 
For large  temperature or  strong disorder, the RFIM exhibits no 
long-range order and behaves like a paramagnet in a field.

Numerically, the nature of this 
phase transition can be studied in many cases conveniently by 
means of exact ground-state (GS) calculations (see below), based on
an established mapping to the graph-theoretical maximum-flow problem. This
allows to treat large system sizes in thermal equilibrium,
in contrast to Monte Carlo simulations.
For Gaussian disorder, the phase transition is of second
order along the full transition line
and can be characterized by critical exponents in the usual way,
like $\nu$, describing the divergence of the correlation length,
and $\alpha$, describing the behavior of the specific heat.
Nevertheless, the RFIM behaves differently
 compared to the standard ferromagnet.
In particular, the hyper-scaling relation $d\nu = 2-\alpha$ has to be 
changed \cite{Grinstein1976} by including a positive parameter $\theta$, 
 yielding 
\begin{equation}
\nu(d - \theta) = 2 - \alpha\,.\label{eq:hyperscaling}
\end{equation}
 In general, the value of
$\theta$ can be obtained directly from the scaling
of domain-wall energies, e.g., induced by changing the boundary
conditions, and is
known as \emph{stiffness exponent}. The study of such domain-wall excitations
was pioneered in the field of spin glasses.\cite{mcmillan1984,bray1984}  Here,
a comprehensive understanding of the nature of the behavior  
of two-dimensional spin glasses could be
obtained,\cite{stiff2d,aspect-ratio2002} 
which turned out to be compatible with the droplet 
picture.\cite{mcmillan1984,bray1987,FisherHuse1986,fisher1988}

According the droplet picture,
the energy scaling of droplet-like excitations should be the same
as for domain walls. For two-dimensional spin glasses, this was recently
confirmed
via using modified GS 
algorithms.\cite{droplets2003,droplets_long2004,pm_droplets2008}
Later on, the value of $\theta$ was
also determined in higher 
dimensions\cite{stiff1999,stiff4d1999,boettcher2004b,%
defect2D2005,boettcher2005}  up to the upper critical dimension, 
in this case via (heuristic) 
ground-state calculations before and after changing the boundary conditions,
respectively.

For the RFIM, which is the subject of this work,
domain-wall studies similar to the spin-glass case, i.e.
based on GS calculations, were performed
for three- and four dimensions,\cite{middleton2002,middleton2002b}
but according our knowledge not in higher dimensions.
Droplet type low- or lowest-energy excitations were only obtained in three
dimensions so far.\cite{fes_rfim2008,Zumsande2009} 
For three dimensions at finite temperature, also free-energy
barrriers where calculated recently.\cite{Fytas2008,Vink2010}

Furthermore, the prediction for the RFIM upper
critical dimension $d_u=6$ was confirmed \cite{AhrensHartmann2011} 
via exhaustive exact GS calculation up to $d=7$. 
This shows that the RFIM can be investigated conveniently by numerical
exact algorithms even close to and above the upper critical dimension. 

Hence, it is the purpose of this work to study domain-wall
and droplet 
excitations  of the RFIM in dimensions $d=5,6,7$ (and for $d=3,4$ for
comparison),
similar to the corresponding $d=2$ studies  for spin glasses, with
the striking difference that for the RFIM an exact polynomial-time 
GS algorithm is available for any dimension, allowing to treat much larger
system sizes of more than five million spins in thermal equilibrium. 
Another difference is that we performed the study for the RFIM  right at the
zero-temperature disorder-critical point (like the previous work in lower
dimensions), since the scaling in the ferromagnetic
and paramagnetic phases should be trivial.
We analyzed energetic and geometric properties of the excitations using 
finite-size scaling. We compare the results of different excitations, which
should
agree according the droplet picture, and verify the above mentioned
 hyper-scaling relation. Also we compare the geometric (fractal) properties
of these excitations.
Treating system sizes up to and above
the upper critical dimension allows us to observe the transition to mean-field
behavior.

To state the model in detail, the RFIM consists of $N$ Ising spins $S_i=\pm 1$ 
located on the sites of a hyper-cubic lattice with periodic boundary conditions
(PBC) 
in all directions. The spins couple to each other and to local net fields. Its
Hamiltonian reads 
\begin{equation}
\mathcal{H}=-J\sum_{\langle i,j\rangle}S_iS_j - \sum_i\left(h \varepsilon_i
\right)S_i\,.
\end{equation}
It has two contributions. The first covers the spin-spin interaction, 
where $J$ is the ferromagnetic coupling constant between two 
adjacent spins and $\langle i,j\rangle$ denotes pairs of
next-neighbored spins. The second part of the Hamiltonian describes the coupling
to local fields $h_i=h \varepsilon_i$,  
The factor $h$ is the disorder strength and $\varepsilon_i$ the quenched
disorder, i.e.  Gaussian distributed with  zero mean and unit width.

The paper is organized as follows: In section \ref{GS} we describe in principle
how the GSs are calculated. 
The next section covers the definition and
use of the different excitations and their theoretical background. 
Then we state
our results in section\ \ref{Results} and finish with our conclusions and 
discussion.

\section{Ground states\label{GS}}
The phase space of the RFIM consists of a ferromagnetic and a
paramagnetic phase. The transition from one phase to the other takes
place at a critical point $P_c=(h_c,T_c)$. The transition can be
triggered varying the temperature $T$ or varying the standard deviation of the
disorder distribution $h$. From Ref.\ \onlinecite{BrayMoore1985}, it is
known via renormalization group calculations, that the critical
behavior of the RFIM is controlled from the zero-temperature fixed
point. Hence, it is possible to focus on  $T=0=\text{const.}$ and
vary $h$ to study the phase transition. Here, we
concentrated on $T=0,\,h\approx h_c$, to study excitations right at 
at the critical point.

At $T=0$ it is possible to calculate exact ground states  in a very
efficient way. Following an approach from Ref. \
\onlinecite{PicardRatliff1975,ogielski1986},
a $d$-dimensional hypercubic realization of the disorder 
$\{h\varepsilon_i\}$ can
 be mapped to a graph with $N+2$ nodes and $(d+2)N+1$ edges with suitable
edge capacities, where $N$ is the number of spins of the RFIM. On this graph a
sophisticated maximum flow/minimum
cut algorithm can be applied.\cite{GoldbergTarjan1988,HartmannRieger2001} The
resulting minimum cut directly correspond to the GS spin configuration 
$\{S_i\}$
of that specific realization of the disorder. For our simulations the
implementation 
of the maximum flow algorithm from the LEDA library\cite{leda1999} 
is used. For the RFIM, the actual runtime of the algorithm increases
only slightly stronger than linear with the number $N$ of spins.
\cite{middleton2001b}
 
\section{Domain walls and droplet excitations\label{num_meth}}

We studied two types of excitations, domain walls and droplets.

\begin{figure}[th]
\begin{center}
\includegraphics[width=0.8\columnwidth]{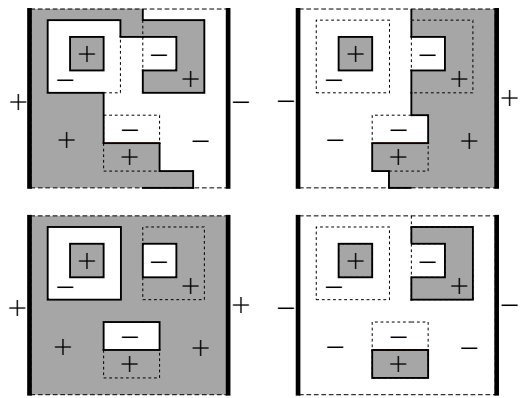}
\caption{Fixing the boundary spins (black bars on the left and right
of every sketch), forces controllable spins to the direction of the
boundary-sign. Areas with $s_i=+1$ are encoded in gray, areas with
spins $s_i=-1$ are white. The $d-1$-dimensional surface between
spins of opposite direction is called  domain wall (black lines).  In
some regions, the random fields may freeze the spins into
boundary-independent stable islands. Islands may include
opposite-directed islands, etc. Hence, islands may also interfere with the
domain wall. The dashed lines signal the border of islands within an
area of equally-oriented spins, and, at the top and the bottom,
the periodic boundary conditions in the remaining $d-1$
directions. \label{fig:boundary_cond}}
\end{center}
\end{figure}

 The domain walls treated in this work separate spin regions which are
 effected from changed (boundary) conditions from unaffected spins.
 Following Ref.\ [\onlinecite{middleton2002}] we forced boundary spins
 along distinct directions, i.e. up ($+$) or down ($-$),  at opposite
boundaries, respectively.  Hence, the PBC are
released in that direction, while they are preserved in the remaining
$d-1$ directions.
We
 calculated the GSs for the four possible combinations ${+-}
 ,\; {-+}, \; {++} , \; {--}$. 
Three types of spin-regions can be distinguished. The
 first type can be flipped changing a single boundary condition,
 i.e. from $++ \to +-$. We call these spin region \emph{strong
 controllable}. The second type is just called \emph{controllable}, if it can
 be flipped by changing both boundary conditions. The third type forms
 fixed, stable \emph{islands}, unaffected of any boundary-condition
 change.  Examples for regions of such spins are shown in
 Fig.\ \ref{fig:boundary_cond}.

\begin{figure}[th]
\begin{center}
\includegraphics[width=0.25\textwidth]{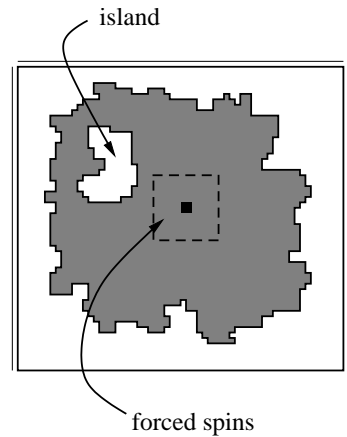}
\caption{Spin-wise difference of the pure GS and  a droplet 
is shown. Unchanged spins are shown as white areas ($s_i - s'_i = 0$),
changed spins are displayed in gray ($|s_i - s'_i| = 2$).\\
The excitation is generated, freezing the spins on the boundaries to their pure
GS orientation, while the spins on a region in the center are  frozen
inverted to their  ground-state orientation (black dot for the
single-spin-induced droplets or dashed square for the bulk-induced droplets).
Under these frozen-spin constraints a new GS is calculated.  
This new configuration is an excitation with respect to the original GS. 
The resulting droplet may spread to an 
arbitrary shape with a fractal surface. It
also may contain islands. \label{fig:droplet_config}}
\end{center}
\end{figure}

We also compare two kinds of droplet excitations of the GS.
  In both cases, the starting point is the GS
calculated for full PBC of a realization.
\begin{itemize}
\item 
The first  type of excitation
is obtained by 
fixing  $(L/3)^d$ spins in the cneter opposite to the
their ground-state orientation, respectively.
 This is inspired by the approach of Ref.\ \onlinecite{Schwarz2009}
for a disordered solid-on-solid model.
This effect can be achieved  conveniently by applying
 strong local fields $\tilde h_i$
in the desired direction for the fixed spins. Also, we fixed the spins on
hyper-planes of the boundary
in parallel to the GS orientation.\cite{FisherHuse1986}
Note that the spins inside the bulk area create a contribution
to the droplet energy via the local fields. To exclude this unwanted
effect, we set the local fields $\varepsilon$ in the bulk region to zero 
for the intitial GS
calculation.
 Via a recalculation of the GS of the modified system, 
this leads to a
large excitation  with respect to the original realization, not including
the boundary, i.e., it is impossible that the fully system flips over. 
 Below, 
 we refer to these excitations as \emph{bulk-induced droplets}, see Fig.\
\ref{fig:droplet_config}. From the definition it is clear that 
these excitations involve $O(L^3)$
spins. Therefore, they come very close to the definition of droplets
in the droplet theory \cite{FisherHuse1986,fisher1988}.
To the knowledge of the authors, such droplets have not been studied
for the RFIM so far.
\item
The other type of excitation, called \emph{single-spin-induced droplets},
 consists  in flipping only the very centered spin
and freeze it anti-parallel to its ground-state orientation,  again
including fixing
boundary spins in parallel to the GS. 
In the same way as for the bulk-induced droplets, 
this is achieved by applying strong local fields.
The droplet created in such a way will include the center spin, but not 
the boundary. Theses singl-spin-induced droplets will be usually
smaller than the bulk-induced droplets. 
Such excitations
have been studied in $d=3$ so far.\cite{Zumsande2009}
\end{itemize}

Hence, for each realization $\{h\varepsilon_i\}$ of the disorder, we
obtained seven different (ground-state) configurations for differt
types of boundary conditions/constraints (PBC,
$(++),\,(--),\,(+-),\,(-+)$, bulk-induced droplets and 
single-spin-induced 
droplets). Technically, to obtain the droplets,
we extracted the absolute differences between the spin configuration
of two or more GSs via linear combinations
of the configurations. Spins with the same properties add to
the same value and form connected clusters. These clusters were obtaineded
using a breadth-fist-search algorithm. In this way,
the calculation of geometric properties of the clusters was very
convenient.  
For the droplets, we measured the domain wall enclosing
the droplet, i.e., the droplet surface.
In the case of different
boundary conditions, we are interested in different definitions of the domain
walls, i.e. separating  uncontrollable spins vs. controllable spins,
strong-controllable spins vs. controllable spins and islands, 
allowing us to measure different
related fractal exponents:

The surfaces are usually not flat or smooth, instead
a any type of disorder-averaged surface $A_\circ$ 
exhibits asymptotically a fractal scaling behavior of
the form
\begin{align}
 A_\circ = c_\circ L^{d_\circ},
\label{eq:fractal}
\end{align}
$d_\circ$ being the corresponding fractal dimension and $c_\circ$ a constant.
Dependent on the different boundary and droplet conditions several fractal
exponents can be deduced. Partially following the definitions of Ref.\
[\onlinecite{middleton2002}] we measured
\begin{enumerate}
 \item[$d_s$:] Surface exponent as dimensionality of the (hyper) surface of
unchanged spins within two GSs of $++$ and $+-$ boundary conditions.
Technically, we calculate  $x_i=|s_i^{++} - s_i^{+-}| = \{0,2\}$ and count the
bonds between the $0$-cluster (unchanged spins) and the $2$-cluster (changed
spins).
 \item[$d_I$:] Incongruent interface exponent as dimensionality of the surface
of unchanged spins within the three GS of $++$, $+-$ and $--$
boundary conditions. This domain-wall does not include parts of any stable
islands. 
In detail, we calculate \mbox{$x_i =  s_i^{--} + s_i^{++}  - 2 s_i^{-+}  + 4 
=\{ 2,4,6 \}$}. It results to $2$ if $s_i$ is strong-controllable.  Spins with
$x_i=4$ belong to a stable island and $x_i=6$, if $s_i$ is just controllable.
For the incongruent boundary the bonds only between the $2$-cluster and
$6$-cluster are counted. 
\item[$d_J$:] Exchange stiffness exponent, defined as the singed sum of broken
bonds, counted positive in the $+-$ and $-+$ configuration and 
negative for $++$
and $--$.
\item[$d_\text{B}$:] fractal exponent of the surface between all flipped and
unflipped spins , i.e., number of bonds between these,
for bulk-induced droplets, including islands.
\item[$d^{(o)}_\text{B}$:] same as above, but excluding islands
\item[$d_\text{1}$:] fractal exponent of the surface between all flipped and
unflipped spins, induced by a single-spin-undiced excitations, 
including islands.
\item[$d^{(o)}_\text{1}$:] same as $d_\text{1}$, but excluding islands.
\end{enumerate}

In the case of
droplet excitations, we measured additionally the disorder-averaged volume
$V$, i.e., the average number of spins in the cluster of droplet spins

Additionally to the geometric properties of domain walls and droplets we are
interested in the stiffness exponent $\theta$, comparing  three different
types of excitations. The first approach is based on 
 the symmetrized stiffness $\Sigma$ defined by Middleton
and Fisher,\cite{middleton2002} i.e., the disorder-averaged
 symmetrized sum of the
boundary-condition dependent energies.
 \begin{align}
  \Sigma \equiv \left\langle  E_{+-} + E_{-+} - E_{++} - E_{--}   
\right\rangle/2\,,
\label{eq:stiffness}
 \end{align}
where $E_{pq}$ is the GS energy for boundary condition 
$pq\in\{++,--,+-,-+\}$,
and
$\langle \cdot\rangle$ denotes the disorder average.

A detailed picture of resulting configurations can be found in Fig.\
\ref{fig:boundary_cond}. Close to criticality, the average stiffness can be
assumed to scale as\cite{middleton2002}
 \begin{align}
 \Sigma(L) \sim  L^\theta\,. \label{eq:stiffness_scaling} 
 \end{align}
$\theta$ denotes the stiffness exponent. 
According the droplet theory, the energy for other types of excitations of 
order of system sizes should scale with the same exponent. Hence,
for the bulk induced droplets, one should be able to observe
\begin{equation}
\Delta E_{\rm B}(L)  \sim L^\theta\,,
\label{eq:bulk:scaling}
\end{equation}
where $\Delta E_{\rm B}$ is the average of the excitation energy of the 
droplet,
i.e., the energy of the droplet configuration for the orginal
value of the fields minus the GS energy for the same original values of
the fields.

Note that the single-spin-induced droplets tend to be small,
i.e., they are not of order of system size. Hence, one cannot directly
measure $\theta$ from the scaling of the droplet energy. Instead,
for the third approach, we 
followe the arguments of Ref.\ [\onlinecite{Zumsande2009}].
Therein, it is shown that the distribution of single-spin-induced 
droplet radii scales with
$p(R)\sim R^{-\theta}$.

\section{\label{Results}Results}
We performed exact ground-state calculations for \begin{tabbing}
\=\phantom{PLATZ!}\=$d=3$:  \=$L=8\dots128$ \=with
\=$11\!\cdot\!10^4$\=$\dots1\!\cdot\!10^5$,\\
\>\>$d=4$:  \>$L=6\dots45$ \>with \>$2\!\cdot\!10^4$
\>$\dots2\!\cdot\!10^5$,\\
\>\>$d=5$: \>$L=6\dots20$ \>with \>$2\!\cdot\!10^3$	\>
$\dots1\!\cdot\!10^4$,\\
\>\>$d=6$: \>$L=6\dots14$ \>with \>$55$ 		
\>$\dots1\!\cdot\!10^4$,\\
\>\>$d=7$: \>$L=4\dots8$ \>with \>$1\!\cdot\!10^3$	\>$\dots1\!\cdot\!10^4$
\end{tabbing} realizations of disorder, where
the largest size exhibit the smallest number of realizations,
respectively. Note that for $d=4$, we studied $L=45$ only
for the bulk-induced droplets, while we did not included results for 
bulk-induced droplets in $d>5$. In general, 
due to computer main memory restrictions we are limited
to system sizes below $5$ million spins.
We start our analysis with the geometric properties of the domain walls.

\subsection{Domain walls}

\begin{center}
\begin{figure}[!ht]
\includegraphics[width=1.1\columnwidth]{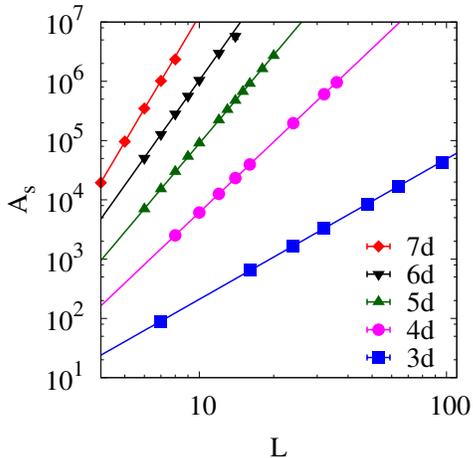}
\caption{(color online) Simple surface of domain walls\label{fig:surface_exp}.
Most error-bars are smaller than the symbol size. The lines
show power-law fits according to Eq.\ (\ref{eq:fractal}).}
\centering
\end{figure}
\end{center}

The surfaces of the 
three different defined boundary-induced domain walls scale with  
plain and very clear power laws. Exemplary, the scaling of the simple 
domain-wall, i.e. $(++)/(+-)$  (yielding the fractal exponent $d_s$)
 is shown in Fig.\ \ref{fig:surface_exp}. 
 Error bars \cite{Hartmann2009}
where obtained as standard error bars from the empirical variance.
The other plots look quite the same with the same precision.  
The scaling exponents $d_s$, $d_J$ and $d_I$ were found with high 
\emph{statistical} accuracy. 
They are stated in Tab.\ \ref{tab:domain_wall_exp}. 
Note that the upper limit for any fractal dimension is $d$, hence the
result for $d_s$ at $d=7$ is an artifact created by the small range
of sizes which is accessible at this high dimension.
Also, the obtained
exponents depend  on the fit range, reflecting
possible systematic correction to the limiting scaling
behavior of Eq.\ (\ref{eq:fractal}).  To estimate such systematic errors,
we have performed fits for different ranges, leading to the
final results which are displayed, c.f. Tab.\ \ref{tab:results}.
Given this accuracy,
the results for $d_J$ may $d_I$ agree or differ, 
in particular in large dimensions, see discussion in Sec.\ \ref{Con},
but $d_s$ differs, comparable
to the previously obtained results\cite{middleton2002,middleton2002b} 
for $d=3,4$.
\begin{table}
\begin{center}
\begin{tabular}{ c  || c | c || c | c|| c | c}
$d$		&$c_s $ 		&$d_s$		& $c_J $ 	& $d_J$	
&$c_I $ 		&$d_I$\\\hline
$3$		&$0.90(3)$	&$2.367(9)$	&$5.9(2)$	& $2.178(8)$
&$0.96(3)$	&$2.28(1)$\\
$4$		&$0.674(8)$	&$3.924(3)$	&$9.0(2)$	& $3.001(7)$
&$1.261(6)$	&$3.231(2)$\\
$5$		&$0.98(2)$	&$4.96(1)$	&$11.7(1)$	& $3.800(4)$
&$1.97(1)$	&$3.925(3)$\\
$6$		&$1.37(9)$	&$5.88(3)$	&$16.2(8)$	& $4.56(2)$
&$3.5(2)$	&$4.51(3)$\\
$7$		&$1.11(9)$	&$7.06(7)$	&$15(1)$	& $5.57(4)$
&$3.3(3)$	&$5.49(5)$
\end{tabular}
\caption{Fit parameters for the scaling of the different surface definitions,
i.e. pure surface, $\Sigma_J$ and the incongruent parts of the domain walls.
\label{tab:domain_wall_exp}}
\end{center}
\end{table}

%

\begin{center}
\begin{figure}[!ht]
\includegraphics[width=1.1\columnwidth]{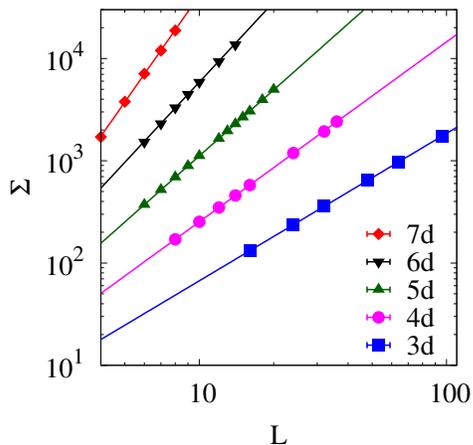}
\caption{(color online) Stiffness, as defined in Eq.
(\ref{eq:stiffness})\label{fig:stiffness}. Most error bars are smaller than the
symbol size. Lines display results of fits to power-laws according
to Eq.\ (\ref{eq:stiffness_scaling}).}
\end{figure}
\end{center}

We continue calculating the stiffness exponents. First, the well
established ansatz according to Eq.\ (\ref{eq:stiffness}) is used for
$d=3,4,5,6,7$. Our data show well behaved power laws in each
dimension, see Fig.\ \ref{fig:stiffness}. The fits of the data points
follow  Eq.\ (\ref{eq:stiffness_scaling}), the parameters are stated
in Tab.\ \ref{tab:stiffness}, in particular the resulting estimate
for the stiffness exponentn $\theta$ is denoted as $\theta_{\rm dw}$.
Varying the ranges of fitted sizes leads to the final estimates,
again stated in Tab.\ \ref{tab:results}.
\begin{table}
\begin{center}
\begin{tabular}{ c  | c | c }
$d$	& 		$a $ 		&  $\theta_{\rm dw}$	\\\hline
$3$	&	$2.42(1)$		&	$1.442(2)$ \\
$4$ &	$4.40(3)$		&	$1.760(3)$ \\
$5$ &	$7.97(5)$		&	$2.146(3)$ \\	
$6$ &	$14.6(5)$		& 	$2.60(1)$  \\
$7$ &	$14.3(6)$		&	$3.45(2)$ 
\end{tabular} 
\caption{Fit parameters for $\theta$ according to Eq.\
(\ref{eq:stiffness_scaling})\label{tab:stiffness}}
\end{center}
\end{table}

\subsection{Droplets}

 Concerning the droplet behavior, we start with the fractal surface properties
of the droplets.

\begin{center}
\begin{figure}[!ht]
\includegraphics[width=1.05\columnwidth]{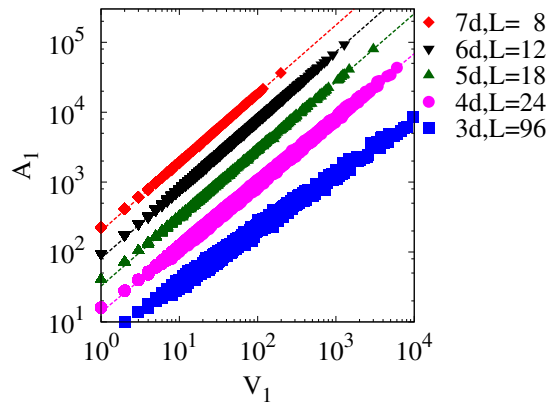}
\caption{(color online) Scatter plot of the enclosing surface,
i.e., not counting islands inside the droplet, of 
single-spin-induced
droplets as function of their volume. The surface is scaled up with a
factor of $k=1,2,4,8,16$ for $d=3,4,5,6,7$ to separate the data visually.
\label{fig:dr_sesurf}}
\centering
\end{figure}
\end{center}

We start by using
 a scatter plot to obtain the relation between surface and volume of 
single-spin-induced
droplets. Theoretically, it should follow a power law of the form
\begin{align}
A\sim V^{d_1/d} \label{eq:AV}.
\end{align}
 The fractal surface exponent with and without stable 
islands can be obtained this way, depending on whether the islands
are included in the calculation of $A$.
And indeed, the results exhibit clear power laws, 
see Fig.\ \ref{fig:dr_sesurf}). 
Additionally, for each dimension the data points for other system
sizes $L$ scatter around the same line, respectively.
 The fit parameters are listed in 
Tab.\ \ref{tab:droplet_wall_exp}, for fixed system sizes.
 The fractal exponents including or ignoring enclosed islands do not 
differ significantly. We have observed slight changes of the value
of $d_1$ when varying the system size $L$ (unless using very small
system sizes where the fitted value of $d_1$ is much smaller). 
Hence, the final values we quote, c.f. 
Tab.\ \ref{tab:results}, are slightly different and involve larger
error bars than the pure statistical error bars.
When approaching
large dimensions $d_s$ gets close to the dimension of the system,
in particular right at the upper critical dimension. The result for
the fractal dimensions is, given the unknown correlations to scaling, in fair
agreement with the results for $d_s$ as shown in Tab.\
\ref{tab:domain_wall_exp}.

\begin{table}
\begin{center}
\begin{tabular}{ c  || r | r || r | r}
$d$		&$c_\text{1} $&$d_\text{1}$&$c^{(o)}_\text{1}$& 
$d^{(o)}_\text{1}$\\\hline
$3$		&$6.07(2)$	&$2.32(1)$	&$6.02(2)$	& $2.33(1)$		\\
$4$		&$6.88(1)$	&$3.70(1)$	&$6.87(1)$	& $3.70(1)$	\\
$5$		&$8.15(3)$	&$4.87(1)$	&$8.51(2)$	& $4.87(1)$		\\
$6$		&$10.13(1)$	&$5.91(1)$	&$10.13(1)$	& $5.91(1)$
\\
$7$		&$12.49(3)$	&$6.88(1)$	&$12.53(3)$	& $6.86(1)$		
\end{tabular}
\caption{Fit parameters for the scaling of the enclosing surface and ragged
surface for single-spin-induced 
droplets, calculated in a large system size with
high sample number, system sizes as in Fig.\ \ref{fig:dr_sesurf}. 
\label{tab:droplet_wall_exp}}
\end{center}
\end{table}

The bulk-induced droplets
have the disadvantage that a smaller effective range
of sizes if accessible: The minimal system size to generate
such a droplet excitation must be $L=6$. In linear direction, one spin
is used to fix the boundary and at least another one is needed at the
very center to form the ``large core''.  Though, only two spin in each
direction are left to form the droplet. In the same way, also for larger
sizes, the volume accessible for the droplets to form is smaller,
leading to stronger finite-size corrections, see below.
 This means, for $d>5$ the range of system sizes ($L\geq6$) 
is too small to observe the leading scaling behavior.  Hence,
we restrict our analysis for the bulk-induced
droplets to $d=3,4$ and $d=5$.

\begin{center}
\begin{figure}[ht]
 \includegraphics[width=1.1\columnwidth]{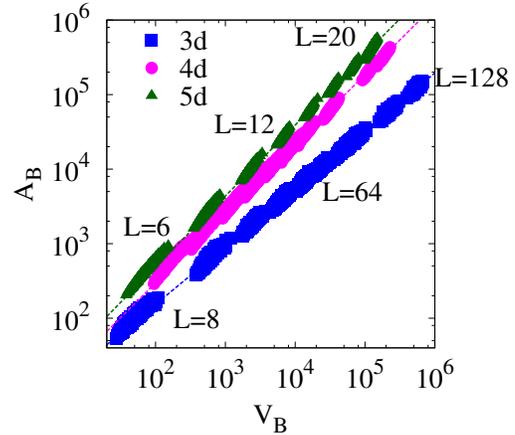}
\caption{(color online) Scatter plot of the enclosing surface of bulk-induced
droplets as function of their volume for $d=3,4,5$. The lower right labels show
system sizes of the $3d$ data, the upper left show those for $5d$. $4d$ is
unlabeled.\label{fig:large_droplet_encl_surface}}
\end{figure}
\end{center}

 In Fig.\ \ref{fig:large_droplet_encl_surface} a scatter-plot of the enclosing
surface as function of the volume can be seen. For the fits the
ansatz according Eq.\ (\ref{eq:AV}) was made,  but using $d_B$ instead
of $d_1$. From the large number of
data points the fit parameters can be obtained, by using the data points
for all system sizs in one fit, with high statistical accuracy.
The
fit parameters can be found in Tab.\ \ref{tab:large_droplet_encl_surface}. 
Given the small ranges of 
system sizes here
and unknown corrections to scaling, the agreement with the results for
the single-spin-induced droplets is fair.

\begin{table}
\begin{center}
\begin{tabular}{ c  || c | c | c|| c | c | c}
$d$		&$c_B $		&$d_B$			&$c_B^{(o)} $ &$d_B^{(o)}$\\\hline
$3$		&$4.300(4)$	&$2.3282(2)$	&$2.761(3) $	&$2.660(2)$\\
$4$		&$4.085(3)$	&$3.7257(2)$	&$4.024(3)$	&$3.7328(2)$\\
$5$		&$6.10(3)$	&$4.758(2)$	&$6.10(3)$	&$4.758(2)$
\end{tabular}
\caption{Fit parameters for the scaling of the enclosing surface and ragged
surface for bulk-induced droplets. For $d>4$, data is not available
in high quality for large-enough system sizes.
\label{tab:large_droplet_encl_surface}}
\end{center}
\end{table}

Systematic finite-size corrections are negelected by this single fit
of the scatter-plot data. Hence, to estimate these corrections,
 leading to the results stated in Tab.\ \ref{tab:results},
we proceeded as follows:
Instead of observing the scatter-plot, the mean of the enclosing
surface can be analyzed as function of the system size. A power law as
$A_{B}(L)=c_B(L-L_0)^{d_B}$ leads to the best fits. The length-scale
correction $L_0$ covers finite-size effects. Of course, an ansatz
using a correction term $~c_BL^{d_B}(1+ c_2L^{d'})$ can be used too,
but the correction term involves one more parameter.
Anyway, fitting $A_{B}(L)$, leads to similar
exponents compared to those as obtained from the scatter plot,
see Tab.\ \ref{tab:large_droplet_encl_surfaceB}.
Again, the results for the different approaches for surface measurement
do not lead to significantly different results.

\begin{table}
\begin{center}
\begin{tabular}{ c  || c | c | c|| c | c | c}
$d$	&$c_B $		&$L_0$		&$d_B$		&$c_B^{(o)} $ &$L_0^{(o)}$		&$d_B^{(o)}$\\\hline
$3$	&$1.37(8)$	&$2.2(1)$	&$2.35(2)$	&$1.02(6) $ &$1.88(9)$		&$2.44(1)$\\
$4$	&$0.49(5)$	&$1.50(9)$	&$3.79(3)$	&$0.48(5)$ &$1.49(9)$		&$3.80(3)$\\
$5$	&$0.8(2)$	&$2.0(1)$	&$4.54(6)$	&$0.8(2)$ &$2.0(1)$		&$4.54(6)$
\end{tabular}
\caption{Fit parameters for the scaling of the enclosing surface and 
ragged surface for bulk-induced droplets when scaling the
surface as a function of system size. For $d=6,7$, a large-enough 
range of system sizes is not available.\label{tab:large_droplet_encl_surfaceB}}
\end{center}
 \end{table}

Next, we discuss energetic properties of droplets, aiming at estimating the 
stiffness exponent $\theta$.

\begin{center}
\begin{figure}[!b]
 \includegraphics[width=1.1\columnwidth]{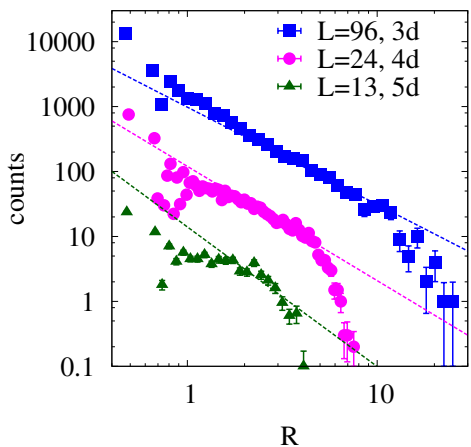}
\caption{(color online) Number of single-spin-induced
 droplets  with a radius in a
specific interval for $d=3,\;L=96$, $d=4,\;L=24$ and $d=4,\;L=13$. The latter
down scaled by a factor $10$ and $20$, respectively. The lines are power laws
using the stiffness exponents.  \label{fig:log_hist_R_small_dr}}
\end{figure}
\end{center}

The singe-spin droplets are known to tend to be very
small.\cite{Zumsande2009} This property prohibits the direct
measurement of the stiffness exponents from the relation of droplet
energy to linear droplet size. Nevertheless, the distribution of
droplet radii $R$ follows\cite{Zumsande2009} approximately a power law
$P(R)\sim R^{-\theta}$, when the data is binned logarithmically.
Hence, we obtained the distribution of droplet radii. For $d=3,4,5$ a
logarithmic binning of the data points is possible. For larger
dimensions, the achievable systems sizes seem to be too small to get a
histogram  of sufficient statistical quality. The histograms can be
seen in Fig.\ \ref{fig:log_hist_R_small_dr}. There exist regions in
which the radii distribution follows $R^{-\theta}$,  when using the
values $\theta_{\rm dw}$ obtained from the domain-wall
measurements. This confirms somehow the results of
Ref. \cite{Zumsande2009} for higher dimensions $d=4,5$.  Nevertheless,
the distribution itself is not sufficient to determine the stiffness
in a meaningful way if the value of $\theta$ was not known from other
sources.

\begin{center}
\begin{figure}[ht]
\includegraphics[width=0.7\columnwidth]{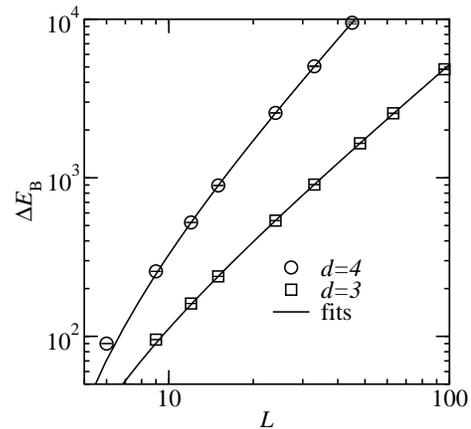}
\caption{Scaling of the excitation energy $\Delta E_{\rm B}$ of bulk-induced
droplets for $d=3$ and $d=4$. Lines show fits to power-laws
$\Delta E_B \sim (L-L_0)^\theta$, with $\theta=1.50$ (3d) and $\theta=1.91$ 
(4d).
\label{fig:larger_droplet_ex_energy}}
\end{figure}
\end{center}

 For the bulk-induced droplets, their length scale is fixed to be
$O(L)$, as assumed in the droplet theory
\cite{FisherHuse1986,fisher1988}. Hence, one could hope that the
scaling of the droplet energy is governed by the (stiffness) exponent
$\theta$. We looked at the scaling of the difference between
the GS energies of the original system and the GS
for  the bulk-induced droplets (calculated with the original
set of random fields, respectively). The resulting 
average bulk droplet energies $\Delta E_{\rm B}$  are shown in 
Fig.\ \ref{fig:larger_droplet_ex_energy} for $d=3,4$.
  A clear curvature is visible,
hence strong finite-size corrections to Eq. (\ref{eq:bulk:scaling}).
This can be explained by the fact that, due to the extensive size
of the bulk area, the effective volume which is accessible for the
droplet to arrange is much smaller. In particular this means, that
for higher dimension $d>4$ the linear sizes $L$ which are accessible
are to small to come even just near to the final scaling
behavior. Hence, we have restricted ourself for the energetic 
properties of the bulk-induced droplets
to dimensions $d=3$ and $4$.
To include corrections to scaling, we fitted the data can by using 
\begin{equation}
\Delta E_B = a(L-L_0)^{\theta_{\rm B}}
\label{eq:stiffness_scaling_corr}\,,
\end{equation}
 see Tab.\
\ref{tab:no_stiffness}. The minimum system size $L_{\min}$ included
in the fit was chosen such that the fitting quality (measured by
the weighted sum of square residuals per degree of freedeom,
denoted as \verb!WSSER/ndf! in {\tt gunplot}) was acceptable, i.e., around 1.
To furthermore estimate systematic corrections, we  also
performed fits for larger values of $L_{\min}$, respectively, leading
to the final results for $\theta_{\rm B}$ 
as stated in Tab.\ \ref{tab:results}. Within error bars, the
values for the droplet stiffness are compatible with the results
obtained for the domain walls. Hence, it appears
that indeed the basic assumptions of the droplet theory is true that
different types of excitations are universally described by the
same exponents.

\begin{table}[ht]
\begin{center}
\begin{tabular}{cccccc}
$d$ &  $L_{\min}$ & $a$ & $L_0$ & $\theta_{\rm B}$  \\\hline
3 & 15 & 5.34(15) & 2.42(10) & 1.501 (7)  \\
4 & 12 & 7.41(18) & 2.75 (5) & 1.91 (1) \\
\end{tabular}
\end{center}
\caption{Fit parameters for the scaling of the droplet energy 
according to Eq.\
(\ref{eq:stiffness_scaling_corr})\label{tab:no_stiffness}. The
fits were performed for a range of system sizes rangeing from
 $L_{\min}$ to the largest system sizes considered here,
respectively.}
\end{table}

 \subsection{Scaling of ground-state energy}

There seems to exist a third alternative way of determining the stiffness 
exponent, not related to externally induced 
domain walls and droplets. For spin glasses
in finite and low dimensions with PBC, 
it has been conjectured\cite{BouchaudKrzakalaMartin2003} and numerically
observed 
for two dimensions \cite{e_scaling2004} that the finite-size
behavior of the (total) GS energy $E_L$ of the unperturbed system
is given by
\begin{equation}
E_L = E_\infty + aL^\theta\,.
\label{eq:e_GS}
\end{equation}
\begin{center}
\begin{figure}[!t]
\includegraphics[width=0.8\columnwidth]{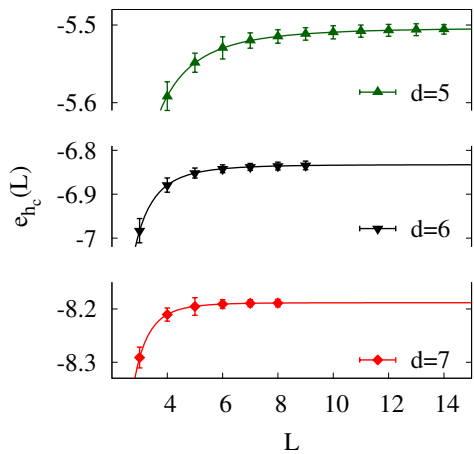}
\caption{(color online) Scaling of the ground-state energy at the 
critical point. The lines show fits to Eq.\ (\ref{eq:e_GS}).
\label{fig:hidden_domainWall_ex_energy}}
\end{figure}
\end{center}
The explanation is that the PBC induce
``hidden'' domain walls (with respect to free boundary conditions),
which dominate the finite-size corrections above higher order contributions.
Hence, we have reanalyzed the data of previous work. 
\cite{HartmannYoung2001,Hartmann2002,AhrensHartmann2011} 
The results of $\theta$ from fitting 
(\ref{eq:e_GS}) to the GS energies can be found in 
Fig.\ \ref{fig:hidden_domainWall_ex_energy}, its exponents are listed 
as $\theta_{\rm E}$ in
Tab.\ \ref{tab:results}. Comparing with the results
for the stiffness exponent $\theta_{\rm dw}$, the assumptions seems
to be indeed true for $d<d_u=6$, maybe also at $d=d_u$ 
(given that we state here only statistical
error bars, hence true error bars are likely larger).  
The assumption of Eq.\ (\ref{eq:e_GS})
is certainly not true above
the upper critical dimension. It seems rather that the exponent for 
GS energy
finite-size correction attains some mean-field value for $d\ge d_u$.
Interestingly, $\theta_{\rm E}$ agrees within error bars
with $\gamma/\nu$ in all dimensions $d=3,\ldots,7$.

\section{\label{Con}Conclusion and Discussion}
To  calculate  ground states of the RFIM numerically, we applied
a well-known mapping to the maximum-flow problem. Using efficient
polynomial-time-running maximum-flow/minimum-cut algorithms,
we were able to study large systems sizes up to $5\times 10^6$ spins
 in exact equilibrium at $T=0$.

Comparing the GS configurations obtained  from different
constraints leads to suitably defined domain-wall and droplet
excitations.
 We obtained  fractal surface exponents  and
the stiffness exponent in $d=3,4,5,6,7$ for these
excitations, see final results in Tab.\ \ref{tab:results}.  The
values for domain walls 
stated in the literature\cite{middleton2002,middleton2002b} for
$d=3,4$ could be recovered.  For all dimensions, different
types of excitations are described, within error bars,
by the same value of the stiffness exponents, as assumed by the
droplet theory. \cite{FisherHuse1986,fisher1988}
In particular, (bulk) droplet excitations have been calculated for the
RFIM for the first time and found to be compatible
with domain walls.
Also, the result for $\theta_\text{dw}$
is according a scaling relation\cite{middleton2002} equal to $d_J-1/\nu$,
which is fulfilled within error bars for all dimensions.
Furthermore, the scaling inequality\cite{Fisher1986,SchwartzSoffer1985} 
$d/2-\beta/\nu\le \theta \le \theta/2$ is fullfilled in all
dimensions.

On the other hand, 
the hyperscaling relation (\ref{eq:hyperscaling}) is fulfilled
within error bars only for $d=3,4,5$, but not for $d=6,7$.
Usually, hyper-scaling relations, involving the dimension $d$,
 are not valid im the mean-field regime, i.e., above the upper
critical dimension. Nevertheless, motivated by our results
for $d=7$, it is compatible with the data if we change the hyperscaling
relation in the following way: We know
from earlier work in high dimensions that above the upper
critical dimension $d_u$  when performing
a data collapse, a rescaling of the linear system size
 according to $L\to L^{d/d_u}$ has
to be performed.\cite{JonesYoung2005}  Since the length rescaling enters  into
scaling relations via the correlation
length\cite{Privman1990,Brezin1982}  $\xi_L/L \sim
\tilde\xi(L^{1/\nu} (h-h_c))$,  the correlation length exponent $\nu$
needs to be replaced with $\nu^*=\nu_\text{MF} d_u/d$
when performing a data collaps. If we transfer this change to
the hyper-scaling relation, it reads $\theta =  d -
(2-\alpha)/\nu^*$. Inserting in $d=7$, $d_u=6$,
$\alpha=0$ and $\nu_\text{MF}=0.5$, this leads to $\theta^{(7)}=3.5$
without any further assumptions. This matches very well to
$\theta^{(7)}=3.5(1)$ as found in this study. 
 On the other hand, a simple rescaling in $d=6$
appears not possible to us, since a replacement $L\to L\ln^{1/6}\!\!L$
was found previously,\cite{AhrensHartmann2011} so rescaled exponents
can not be obtained in a simple way,  hence a different kind of
hyperscaling seems to be necessary here.

Also the fractal properties of the droplets and domain walls found
for lower dimensions extent to larger dimensions. In particular,
the ``geometrically'' measured values $d_s$, $d_1$ and $d_{\rm B}$
agree for all dimensions within error bars. Finally,
the values for $d_{I}$ and $d_{J}$, measuring different fractal properties
of the domain walls, may or may  not be equivalent, as already discussed
in previously\cite{middleton2002,middleton2002b} for 
the cases $d=3$ and $d=4$,
were also a possible scaling relation
$d_{I}=d_{J}+\beta/\nu$ was presented.
For $d\ge 6$ this relation is clearly not fullfiled here, 
which may be also due to the 
failing of hyperscaling at and above the upper critical dimension.

\begin{table}
\centering
\small
\begin{tabular}{c  | l l l l l }
 			&  $d=3$ 	& $d=4$ 		& $d=5$		
& $d=6$ 		&$d=7$ 		\\ \hline
$h_c$		& $ 2.28(1)$	& $4.18(1)$		& $6.02(2)$	&
$7.78(1)$	&$9.48(3)$\\
$\beta$		& $ 0.017(5) $& $0.13(5) $		& $0.25(1)$	
& $0.50(5)$		& $0.50(5)$	\\
$\gamma$	& $ 1.98(7)$	& $1.57(10) $	& $1.3(1)$		&
$1.07(5)$		&$1.0(2)$	\\
$\alpha$		& $0$  		& $ 0 $			& $ 0 $		
& $ 0 $			& $0$		\\
$\nu$ 		& $ 1.37(9)$	& $0.78(10)$		& $0.60(3)$	&
$0.50(5)$		& $0.47(5)$	\\
$\tilde\theta=\gamma/\nu$ & $1.4(2) $ & $2.0(4)$  &$2.3(3)$ 		& $2.1(5)$	
&$2.2(5) $ \\\hline\hline
$\theta_\text{\rm E}$&$1.49(3)$	& $1.81(6)$& $2.03(2)$&
$2.42(2)$	& $2.42(2)$\\
$\theta_\text{dw}$&$1.44(2)$	&$1.75(2)$	&$2.15(1)$	
&$2.60(2)$		&$3.5(1)$\\
$d_J\!-\!1/\nu$&$1.4(1)$   	& $1.65(15)$ 		&$2.1(1)$	
&$2.5(2)$		&$3.5(1)$\\
$\theta_{\rm B}$ 	&$1.51(2)$	&$1.8(1)$		&---	
&---				&---\\ \hline
$d_s$		&$2.37(2)$	&$3.92(6)$ 		&$4.9(2)$	
&$5.8(8)$		&$6.9(2)$\\
$d_1$		&$2.32(1)$	&$3.73(10) $		&$4.9(1) $	
&$5.9(1) $		&$6.9(1)$\\
$d_B$		&$2.35(2)$	&$3.79(4) $		&$4.6(1)$	
&---				&---\\
$d_J$		&$2.14(3)$	&$2.93(2)$ 		&$3.79(2)$	
&$4.54(2)$		&$5.5(1)$\\
$d_I$		&$2.25(2)$	&$3.23(1)$		&$3.92(1)$	
&$4.5(2)$		&$5.4(2)$		
 \end{tabular}
\caption{Previous 
results\cite{middleton2002,middleton2002b,AhrensHartmann2011}  
above the big line and new, final results below. 
The upper half of the final results contains different
estimates for the stiffness exponent, while the bottom part
contains measured for the fractal properties of domain walls.
Note that the values stated here differ usually 
from the values given in the lists of fit parameters 
since they the former include systematical errors,
which were estimated by varying the range of fitted system sizes
and observing the change of the resulting parameters.
\label{tab:results}}
\end{table}

\section{\label{Ack}Acknowledgments}
We thank K. Binder, S. Boettcher, A. Hucht, O. Melchert,
M. M\'ezard, U. Nowak, H. Rieger, 
  D. Sherrington, and D. Stauffer for useful discussions.
The calculations were carried out on \textbf{GOLEM} (\textbf{G}ro\ss rechner
\textbf{OL}denburg f\"ur \textbf{E}xplizit \textbf{M}ultidisziplin\"are
\textbf{F}orschung) and the \textbf{HERO} (\textbf{H}igh-\textbf{E}nd Computing
\textbf{R}esource \textbf{O}ldenburg) at the University of Oldenburg.
\newpage
\bibliographystyle{apsrev}
\bibliography{Literatur}

\begin{thebibliography}{48}
\expandafter\ifx\csname natexlab\endcsname\relax\def\natexlab#1{#1}\fi
\expandafter\ifx\csname bibnamefont\endcsname\relax
  \def\bibnamefont#1{#1}\fi
\expandafter\ifx\csname bibfnamefont\endcsname\relax
  \def\bibfnamefont#1{#1}\fi
\expandafter\ifx\csname citenamefont\endcsname\relax
  \def\citenamefont#1{#1}\fi
\expandafter\ifx\csname url\endcsname\relax
  \def\url#1{\texttt{#1}}\fi
\expandafter\ifx\csname urlprefix\endcsname\relax\def\urlprefix{URL }\fi
\providecommand{\bibinfo}[2]{#2}
\providecommand{\eprint}[2][]{\url{#2}}

\bibitem[{\citenamefont{Bricmont and Kupiainen}(1987)}]{bricmont1987}
\bibinfo{author}{\bibfnamefont{J.}~\bibnamefont{Bricmont}} \bibnamefont{and}
  \bibinfo{author}{\bibfnamefont{A.}~\bibnamefont{Kupiainen}},
  \bibinfo{journal}{Phys. Rev. Lett.} \textbf{\bibinfo{volume}{59}},
  \bibinfo{pages}{1829} (\bibinfo{year}{1987}).

\bibitem[{\citenamefont{Gofman et~al.}(1993)\citenamefont{Gofman, Adler,
  Aharony, Harris, and Schwartz}}]{GofmanAdlerAharonyHarrisSchwartz1993}
\bibinfo{author}{\bibfnamefont{M.}~\bibnamefont{Gofman}},
  \bibinfo{author}{\bibfnamefont{J.}~\bibnamefont{Adler}},
  \bibinfo{author}{\bibfnamefont{A.}~\bibnamefont{Aharony}},
  \bibinfo{author}{\bibfnamefont{A.~B.} \bibnamefont{Harris}},
  \bibnamefont{and} \bibinfo{author}{\bibfnamefont{M.}~\bibnamefont{Schwartz}},
  \bibinfo{journal}{Phys. Rev. Lett.} \textbf{\bibinfo{volume}{71}},
  \bibinfo{pages}{1569} (\bibinfo{year}{1993}).

\bibitem[{\citenamefont{Rieger}(1995)}]{Rieger1995}
\bibinfo{author}{\bibfnamefont{H.}~\bibnamefont{Rieger}},
  \bibinfo{journal}{Phys. Rev. B} \textbf{\bibinfo{volume}{52}},
  \bibinfo{pages}{6659} (\bibinfo{year}{1995}).

\bibitem[{\citenamefont{Nowak et~al.}(1998)\citenamefont{Nowak, Usadel, and
  Esser}}]{Nowak1998}
\bibinfo{author}{\bibfnamefont{U.}~\bibnamefont{Nowak}},
  \bibinfo{author}{\bibfnamefont{K.~D.} \bibnamefont{Usadel}},
  \bibnamefont{and} \bibinfo{author}{\bibfnamefont{J.}~\bibnamefont{Esser}},
  \bibinfo{journal}{Physica A: Statistical and Theoretical Physics}
  \textbf{\bibinfo{volume}{250}}, \bibinfo{pages}{1 } (\bibinfo{year}{1998}).

\bibitem[{\citenamefont{Hartmann and Nowak}(1999)}]{art_uli1999}
\bibinfo{author}{\bibfnamefont{A.~K.} \bibnamefont{Hartmann}} \bibnamefont{and}
  \bibinfo{author}{\bibfnamefont{U.}~\bibnamefont{Nowak}},
  \bibinfo{journal}{Eur. Phys. J. B} \textbf{\bibinfo{volume}{7}},
  \bibinfo{pages}{105} (\bibinfo{year}{1999}).

\bibitem[{\citenamefont{Hartmann and
  Young}(2001{\natexlab{a}})}]{HartmannYoung2001}
\bibinfo{author}{\bibfnamefont{A.~K.} \bibnamefont{Hartmann}} \bibnamefont{and}
  \bibinfo{author}{\bibfnamefont{A.~P.} \bibnamefont{Young}},
  \bibinfo{journal}{Phys. Rev. B} \textbf{\bibinfo{volume}{64}},
  \bibinfo{pages}{214419} (\bibinfo{year}{2001}{\natexlab{a}}).

\bibitem[{\citenamefont{Middleton and Fisher}(2002)}]{middleton2002}
\bibinfo{author}{\bibfnamefont{A.~A.} \bibnamefont{Middleton}}
  \bibnamefont{and} \bibinfo{author}{\bibfnamefont{D.~S.}
  \bibnamefont{Fisher}}, \bibinfo{journal}{Phys. Rev. B}
  \textbf{\bibinfo{volume}{65}}, \bibinfo{pages}{134411}
  (\bibinfo{year}{2002}).

\bibitem[{\citenamefont{Frontera and Vives}(2002)}]{frontera2002}
\bibinfo{author}{\bibfnamefont{C.}~\bibnamefont{Frontera}} \bibnamefont{and}
  \bibinfo{author}{\bibfnamefont{E.}~\bibnamefont{Vives}},
  \bibinfo{journal}{Computer Physics Communications}
  \textbf{\bibinfo{volume}{147}}, \bibinfo{pages}{455 } (\bibinfo{year}{2002}).

\bibitem[{\citenamefont{Sepp\"al\"a et~al.}(2002)\citenamefont{Sepp\"al\"a,
  Pulkkinen, and Alava}}]{seppala2002}
\bibinfo{author}{\bibfnamefont{E.~T.} \bibnamefont{Sepp\"al\"a}},
  \bibinfo{author}{\bibfnamefont{A.~M.} \bibnamefont{Pulkkinen}},
  \bibnamefont{and} \bibinfo{author}{\bibfnamefont{M.~J.} \bibnamefont{Alava}},
  \bibinfo{journal}{Phys. Rev. B} \textbf{\bibinfo{volume}{66}},
  \bibinfo{pages}{144403} (\bibinfo{year}{2002}).

\bibitem[{\citenamefont{Hartmann}(2002)}]{Hartmann2002}
\bibinfo{author}{\bibfnamefont{A.~K.} \bibnamefont{Hartmann}},
  \bibinfo{journal}{Phys. Rev. B} \textbf{\bibinfo{volume}{65}},
  \bibinfo{pages}{174427} (\bibinfo{year}{2002}).

\bibitem[{\citenamefont{Middleton}(2002)}]{middleton2002b}
\bibinfo{author}{\bibfnamefont{A.~A.} \bibnamefont{Middleton}},
  \bibinfo{journal}{preprint arXiv:cond-mat/0208182}  (\bibinfo{year}{2002}).

\bibitem[{\citenamefont{Zumsande et~al.}(2008)\citenamefont{Zumsande, Alava,
  and Hartmann}}]{fes_rfim2008}
\bibinfo{author}{\bibfnamefont{M.}~\bibnamefont{Zumsande}},
  \bibinfo{author}{\bibfnamefont{M.~J.} \bibnamefont{Alava}}, \bibnamefont{and}
  \bibinfo{author}{\bibfnamefont{A.~K.} \bibnamefont{Hartmann}},
  \bibinfo{journal}{J. Stat. Mech.} p. \bibinfo{pages}{P02012}
  (\bibinfo{year}{2008}).

\bibitem[{\citenamefont{Ahrens and Hartmann}(2011)}]{AhrensHartmann2011}
\bibinfo{author}{\bibfnamefont{B.}~\bibnamefont{Ahrens}} \bibnamefont{and}
  \bibinfo{author}{\bibfnamefont{A.~K.} \bibnamefont{Hartmann}},
  \bibinfo{journal}{Phys. Rev. B} \textbf{\bibinfo{volume}{83}},
  \bibinfo{pages}{014205} (\bibinfo{year}{2011}).

\bibitem[{\citenamefont{Grinstein}(1976)}]{Grinstein1976}
\bibinfo{author}{\bibfnamefont{G.}~\bibnamefont{Grinstein}},
  \bibinfo{journal}{Phys. Rev. Lett.} \textbf{\bibinfo{volume}{37}},
  \bibinfo{pages}{944} (\bibinfo{year}{1976}).

\bibitem[{\citenamefont{McMillan}(1984)}]{mcmillan1984}
\bibinfo{author}{\bibfnamefont{W.~L.} \bibnamefont{McMillan}},
  \bibinfo{journal}{J. Phys. C} \textbf{\bibinfo{volume}{17}},
  \bibinfo{pages}{3179} (\bibinfo{year}{1984}).

\bibitem[{\citenamefont{Bray and Moore}(1984)}]{bray1984}
\bibinfo{author}{\bibfnamefont{A.~J.} \bibnamefont{Bray}} \bibnamefont{and}
  \bibinfo{author}{\bibfnamefont{M.~A.} \bibnamefont{Moore}},
  \bibinfo{journal}{J. Phys. C} \textbf{\bibinfo{volume}{17}}
  (\bibinfo{year}{1984}).

\bibitem[{\citenamefont{Hartmann and Young}(2001{\natexlab{b}})}]{stiff2d}
\bibinfo{author}{\bibfnamefont{A.~K.} \bibnamefont{Hartmann}} \bibnamefont{and}
  \bibinfo{author}{\bibfnamefont{A.~P.} \bibnamefont{Young}},
  \bibinfo{journal}{Phys. Rev B} \textbf{\bibinfo{volume}{64}},
  \bibinfo{pages}{180404} (\bibinfo{year}{2001}{\natexlab{b}}).

\bibitem[{\citenamefont{Hartmann et~al.}(2002)\citenamefont{Hartmann, Bray,
  Carter, Moore, and Young}}]{aspect-ratio2002}
\bibinfo{author}{\bibfnamefont{A.~K.} \bibnamefont{Hartmann}},
  \bibinfo{author}{\bibfnamefont{A.~J.} \bibnamefont{Bray}},
  \bibinfo{author}{\bibfnamefont{A.~C.} \bibnamefont{Carter}},
  \bibinfo{author}{\bibfnamefont{M.~A.} \bibnamefont{Moore}}, \bibnamefont{and}
  \bibinfo{author}{\bibfnamefont{A.~P.} \bibnamefont{Young}},
  \bibinfo{journal}{Phys. Rev. B} \textbf{\bibinfo{volume}{66}},
  \bibinfo{pages}{224401} (\bibinfo{year}{2002}).

\bibitem[{\citenamefont{Bray and Moore}(1987)}]{bray1987}
\bibinfo{author}{\bibfnamefont{A.~J.} \bibnamefont{Bray}} \bibnamefont{and}
  \bibinfo{author}{\bibfnamefont{M.~A.} \bibnamefont{Moore}}, in
  \emph{\bibinfo{booktitle}{Heidelberg Colloquium on Glassy Dynamics}}, edited
  by \bibinfo{editor}{\bibfnamefont{J.~L.} \bibnamefont{van Hemmen}}
  \bibnamefont{and}
  \bibinfo{editor}{\bibfnamefont{I.}~\bibnamefont{Morgenstern}}
  (\bibinfo{publisher}{Springer}, \bibinfo{address}{Berlin},
  \bibinfo{year}{1987}), p. \bibinfo{pages}{121}.

\bibitem[{\citenamefont{Fisher and Huse}(1986)}]{FisherHuse1986}
\bibinfo{author}{\bibfnamefont{D.~S.} \bibnamefont{Fisher}} \bibnamefont{and}
  \bibinfo{author}{\bibfnamefont{D.~A.} \bibnamefont{Huse}},
  \bibinfo{journal}{Phys. Rev. Lett.} \textbf{\bibinfo{volume}{56}},
  \bibinfo{pages}{1601} (\bibinfo{year}{1986}),
  \urlprefix\url{http://link.aps.org/doi/10.1103/PhysRevLett.56.1601}.

\bibitem[{\citenamefont{Fisher and Huse}(1988)}]{fisher1988}
\bibinfo{author}{\bibfnamefont{D.~S.} \bibnamefont{Fisher}} \bibnamefont{and}
  \bibinfo{author}{\bibfnamefont{D.~A.} \bibnamefont{Huse}},
  \bibinfo{journal}{Phys. Rev. B} \textbf{\bibinfo{volume}{38}},
  \bibinfo{pages}{386} (\bibinfo{year}{1988}).

\bibitem[{\citenamefont{Hartmann and Moore}(2003)}]{droplets2003}
\bibinfo{author}{\bibfnamefont{A.~K.} \bibnamefont{Hartmann}} \bibnamefont{and}
  \bibinfo{author}{\bibfnamefont{M.~A.} \bibnamefont{Moore}},
  \bibinfo{journal}{Phys. Rev. Lett.} \textbf{\bibinfo{volume}{90}},
  \bibinfo{pages}{127201} (\bibinfo{year}{2003}).

\bibitem[{\citenamefont{Hartmann and Moore}(2004)}]{droplets_long2004}
\bibinfo{author}{\bibfnamefont{A.~K.} \bibnamefont{Hartmann}} \bibnamefont{and}
  \bibinfo{author}{\bibfnamefont{M.~A.} \bibnamefont{Moore}},
  \bibinfo{journal}{Phys. Rev. B} \textbf{\bibinfo{volume}{69}},
  \bibinfo{pages}{104409} (\bibinfo{year}{2004}).

\bibitem[{\citenamefont{Hartmann}(2008)}]{pm_droplets2008}
\bibinfo{author}{\bibfnamefont{A.~K.} \bibnamefont{Hartmann}},
  \bibinfo{journal}{Phys. Rev. B} \textbf{\bibinfo{volume}{77}},
  \bibinfo{pages}{144418} (\bibinfo{year}{2008}).

\bibitem[{\citenamefont{Hartmann}(1999{\natexlab{a}})}]{stiff1999}
\bibinfo{author}{\bibfnamefont{A.~K.} \bibnamefont{Hartmann}},
  \bibinfo{journal}{Phys. Rev. E} \textbf{\bibinfo{volume}{59}},
  \bibinfo{pages}{84} (\bibinfo{year}{1999}{\natexlab{a}}).

\bibitem[{\citenamefont{Hartmann}(1999{\natexlab{b}})}]{stiff4d1999}
\bibinfo{author}{\bibfnamefont{A.~K.} \bibnamefont{Hartmann}},
  \bibinfo{journal}{Phys. Rev. E} \textbf{\bibinfo{volume}{60}},
  \bibinfo{pages}{5135} (\bibinfo{year}{1999}{\natexlab{b}}).

\bibitem[{\citenamefont{Boettcher}(2004)}]{boettcher2004b}
\bibinfo{author}{\bibfnamefont{S.}~\bibnamefont{Boettcher}},
  \bibinfo{journal}{Eur. Phys. J. B} \textbf{\bibinfo{volume}{38}},
  \bibinfo{pages}{83} (\bibinfo{year}{2004}), ISSN \bibinfo{issn}{1434-6028},
  \bibinfo{note}{10.1140/epjb/e2004-00102-5},
  \urlprefix\url{http://dx.doi.org/10.1140/epjb/e2004-00102-5}.

\bibitem[{\citenamefont{Boettcher and Hartmann}(2005)}]{defect2D2005}
\bibinfo{author}{\bibfnamefont{S.}~\bibnamefont{Boettcher}} \bibnamefont{and}
  \bibinfo{author}{\bibfnamefont{A.~K.} \bibnamefont{Hartmann}},
  \bibinfo{journal}{Phys. Rev. B} \textbf{\bibinfo{volume}{72}},
  \bibinfo{pages}{014429} (\bibinfo{year}{2005}).

\bibitem[{\citenamefont{Boettcher}(2005)}]{boettcher2005}
\bibinfo{author}{\bibfnamefont{S.}~\bibnamefont{Boettcher}},
  \bibinfo{journal}{Phys. Rev. Lett.} \textbf{\bibinfo{volume}{95}},
  \bibinfo{pages}{197205} (\bibinfo{year}{2005}).

\bibitem[{\citenamefont{{Zumsande, M.} and {Hartmann, A.
  K.}}(2009)}]{Zumsande2009}
\bibinfo{author}{\bibnamefont{{Zumsande, M.}}} \bibnamefont{and}
  \bibinfo{author}{\bibnamefont{{Hartmann, A. K.}}}, \bibinfo{journal}{Eur.
  Phys. J. B} \textbf{\bibinfo{volume}{72}}, \bibinfo{pages}{619}
  (\bibinfo{year}{2009}),
  \urlprefix\url{http://dx.doi.org/10.1140/epjb/e2009-00410-2}.

\bibitem[{\citenamefont{Fytas et~al.}(2008)\citenamefont{Fytas, Malakis, and
  Eftaxias}}]{Fytas2008}
\bibinfo{author}{\bibfnamefont{N.~G.} \bibnamefont{Fytas}},
  \bibinfo{author}{\bibfnamefont{A.}~\bibnamefont{Malakis}}, \bibnamefont{and}
  \bibinfo{author}{\bibfnamefont{K.}~\bibnamefont{Eftaxias}},
  \bibinfo{journal}{Journal of Statistical Mechanics: Theory and Experiment}
  \textbf{\bibinfo{volume}{2008}}, \bibinfo{pages}{P03015}
  (\bibinfo{year}{2008}),
  \urlprefix\url{http://stacks.iop.org/1742-5468/2008/i=03/a=P03015}.

\bibitem[{\citenamefont{Vink et~al.}(2010)\citenamefont{Vink, Fischer, and
  Binder}}]{Vink2010}
\bibinfo{author}{\bibfnamefont{R.~L.~C.} \bibnamefont{Vink}},
  \bibinfo{author}{\bibfnamefont{T.}~\bibnamefont{Fischer}}, \bibnamefont{and}
  \bibinfo{author}{\bibfnamefont{K.}~\bibnamefont{Binder}},
  \bibinfo{journal}{Phys. Rev. E} \textbf{\bibinfo{volume}{82}},
  \bibinfo{pages}{051134} (\bibinfo{year}{2010}),
  \urlprefix\url{http://link.aps.org/doi/10.1103/PhysRevE.82.051134}.

\bibitem[{\citenamefont{Bray and Moore}(1985)}]{BrayMoore1985}
\bibinfo{author}{\bibfnamefont{A.~J.} \bibnamefont{Bray}} \bibnamefont{and}
  \bibinfo{author}{\bibfnamefont{M.~A.} \bibnamefont{Moore}},
  \bibinfo{journal}{J. Phys. C: Solid State Phys.}
  \textbf{\bibinfo{volume}{18}}, \bibinfo{pages}{927} (\bibinfo{year}{1985}).

\bibitem[{\citenamefont{Picard and Ratliff}(1975)}]{PicardRatliff1975}
\bibinfo{author}{\bibfnamefont{J.~C.} \bibnamefont{Picard}} \bibnamefont{and}
  \bibinfo{author}{\bibfnamefont{H.~D.} \bibnamefont{Ratliff}},
  \bibinfo{journal}{Networks} \textbf{\bibinfo{volume}{5}},
  \bibinfo{pages}{357} (\bibinfo{year}{1975}).

\bibitem[{\citenamefont{Ogielski}(1986)}]{ogielski1986}
\bibinfo{author}{\bibfnamefont{A.~T.} \bibnamefont{Ogielski}},
  \bibinfo{journal}{Phys. Rev. Lett.} \textbf{\bibinfo{volume}{57}},
  \bibinfo{pages}{1251} (\bibinfo{year}{1986}).

\bibitem[{\citenamefont{Goldberg and Tarjan}(1988)}]{GoldbergTarjan1988}
\bibinfo{author}{\bibfnamefont{A.~V.} \bibnamefont{Goldberg}} \bibnamefont{and}
  \bibinfo{author}{\bibfnamefont{R.~E.} \bibnamefont{Tarjan}},
  \bibinfo{journal}{J. ACM} \textbf{\bibinfo{volume}{35}}, \bibinfo{pages}{921}
  (\bibinfo{year}{1988}), ISSN \bibinfo{issn}{0004-5411}.

\bibitem[{\citenamefont{Hartmann and Rieger}(2001)}]{HartmannRieger2001}
\bibinfo{author}{\bibfnamefont{A.~K.} \bibnamefont{Hartmann}} \bibnamefont{and}
  \bibinfo{author}{\bibfnamefont{H.}~\bibnamefont{Rieger}},
  \emph{\bibinfo{title}{Optimization Algorithms in Physics}}
  (\bibinfo{publisher}{Wiley-VCH, Berlin}, \bibinfo{year}{2001}), ISBN
  \bibinfo{isbn}{978-3-527-40307-3}.

\bibitem[{\citenamefont{Mehlhorn and N\"aher}(1999)}]{leda1999}
\bibinfo{author}{\bibfnamefont{K.}~\bibnamefont{Mehlhorn}} \bibnamefont{and}
  \bibinfo{author}{\bibfnamefont{S.}~\bibnamefont{N\"aher}},
  \emph{\bibinfo{title}{The LEDA Platform of Combinatorial and Geometric
  Computing}} (\bibinfo{publisher}{Cambridge University Press},
  \bibinfo{address}{Cambridge}, \bibinfo{year}{1999}),
  \urlprefix\url{http://www.algorithmic-solutions.de}.

\bibitem[{\citenamefont{Middleton}(2001)}]{middleton2001b}
\bibinfo{author}{\bibfnamefont{A.~A.} \bibnamefont{Middleton}},
  \bibinfo{journal}{Phys. Rev. Lett.} \textbf{\bibinfo{volume}{88}},
  \bibinfo{pages}{017202} (\bibinfo{year}{2001}),
  \urlprefix\url{http://link.aps.org/doi/10.1103/PhysRevLett.88.017202}.

\bibitem[{\citenamefont{Schwarz et~al.}(2009)\citenamefont{Schwarz,
  Karrenbauer, Schehr, and Rieger}}]{Schwarz2009}
\bibinfo{author}{\bibfnamefont{K.}~\bibnamefont{Schwarz}},
  \bibinfo{author}{\bibfnamefont{A.}~\bibnamefont{Karrenbauer}},
  \bibinfo{author}{\bibfnamefont{G.}~\bibnamefont{Schehr}}, \bibnamefont{and}
  \bibinfo{author}{\bibfnamefont{H.}~\bibnamefont{Rieger}},
  \bibinfo{journal}{Journal of Statistical Mechanics: Theory and Experiment}
  \textbf{\bibinfo{volume}{2009}}, \bibinfo{pages}{P08022}
  (\bibinfo{year}{2009}),
  \urlprefix\url{http://stacks.iop.org/1742-5468/2009/i=08/a=P08022}.

\bibitem[{\citenamefont{Hartmann}(2009)}]{Hartmann2009}
\bibinfo{author}{\bibfnamefont{A.~K.} \bibnamefont{Hartmann}},
  \emph{\bibinfo{title}{A Practical Guide To Computer Si\-mu\-la\-tion}}
  (\bibinfo{publisher}{World Scientific Publishing Company},
  \bibinfo{year}{2009}), ISBN \bibinfo{isbn}{978-9812834157}.

\bibitem[{\citenamefont{Bouchaud et~al.}(2003)\citenamefont{Bouchaud, Krzakala,
  and Martin}}]{BouchaudKrzakalaMartin2003}
\bibinfo{author}{\bibfnamefont{J.-P.} \bibnamefont{Bouchaud}},
  \bibinfo{author}{\bibfnamefont{F.}~\bibnamefont{Krzakala}}, \bibnamefont{and}
  \bibinfo{author}{\bibfnamefont{O.~C.} \bibnamefont{Martin}},
  \bibinfo{journal}{Phys. Rev. B} \textbf{\bibinfo{volume}{68}},
  \bibinfo{pages}{224404} (\bibinfo{year}{2003}).

\bibitem[{\citenamefont{Campbell et~al.}(2004)\citenamefont{Campbell, Hartmann,
  and Katzgraber}}]{e_scaling2004}
\bibinfo{author}{\bibfnamefont{I.~A.} \bibnamefont{Campbell}},
  \bibinfo{author}{\bibfnamefont{A.~K.} \bibnamefont{Hartmann}},
  \bibnamefont{and} \bibinfo{author}{\bibfnamefont{H.~G.}
  \bibnamefont{Katzgraber}}, \bibinfo{journal}{Phys. Rev. B}
  \textbf{\bibinfo{volume}{70}}, \bibinfo{pages}{054429}
  (\bibinfo{year}{2004}).

\bibitem[{\citenamefont{Fisher}(1986)}]{Fisher1986}
\bibinfo{author}{\bibfnamefont{D.~S.} \bibnamefont{Fisher}},
  \bibinfo{journal}{Phys. Rev. Lett.} \textbf{\bibinfo{volume}{56}},
  \bibinfo{pages}{416} (\bibinfo{year}{1986}).

\bibitem[{\citenamefont{Schwartz and Soffer}(1985)}]{SchwartzSoffer1985}
\bibinfo{author}{\bibfnamefont{M.}~\bibnamefont{Schwartz}} \bibnamefont{and}
  \bibinfo{author}{\bibfnamefont{A.}~\bibnamefont{Soffer}},
  \bibinfo{journal}{Phys. Rev. Lett.} \textbf{\bibinfo{volume}{55}},
  \bibinfo{pages}{2499} (\bibinfo{year}{1985}).

\bibitem[{\citenamefont{Jones and Young}(2005)}]{JonesYoung2005}
\bibinfo{author}{\bibfnamefont{J.~L.} \bibnamefont{Jones}} \bibnamefont{and}
  \bibinfo{author}{\bibfnamefont{A.~P.} \bibnamefont{Young}},
  \bibinfo{journal}{Phys. Rev. B} \textbf{\bibinfo{volume}{71}},
  \bibinfo{pages}{174438} (\bibinfo{year}{2005}).

\bibitem[{\citenamefont{Privman}(1990)}]{Privman1990}
\bibinfo{author}{\bibfnamefont{V.~E.} \bibnamefont{Privman}},
  \emph{\bibinfo{title}{Finite Size Scaling and Numerical Simulation of
  Statistical Systems}} (\bibinfo{publisher}{World Scientific Publishing
  Company}, \bibinfo{year}{1990}), ISBN \bibinfo{isbn}{978-9810237967}.

\bibitem[{\citenamefont{Br\'ezin}(1982)}]{Brezin1982}
\bibinfo{author}{\bibfnamefont{E.}~\bibnamefont{Br\'ezin}},
  \bibinfo{journal}{J. Phys. France} \textbf{\bibinfo{volume}{43}},
  \bibinfo{pages}{15} (\bibinfo{year}{1982}).

\end{thebibliography}
\end{document}